\documentclass{article}

\usepackage{arxiv}

\usepackage[utf8]{inputenc} 
\usepackage[T1]{fontenc}    
\usepackage{hyperref}       
\usepackage{url}            
\usepackage{booktabs}       
\usepackage{amsfonts}       
\usepackage{amsmath}
\usepackage{amssymb}
\usepackage{nicefrac}       
\usepackage{microtype}      
\usepackage{lipsum}
\usepackage{graphicx}
\usepackage{dsfont} 

\usepackage{makecell}
\newcounter{magicrownumbers}
\newcommand\rownumber{\stepcounter{magicrownumbers}\arabic{magicrownumbers}}

\title{Constrained Attractor Selection Using Deep Reinforcement Learning}

\author{
	Xue-She Wang, James D. Turner \& Brian P. Mann \\
	Dynamical Systems Research Laboratory \\
	Department of Mechanical Engineering \& Materials Science \\
	Duke University \\
	Durham, NC 27708, USA \\
	\texttt{xueshe.wang@duke.edu} \\
}

\begin{document}
\maketitle

\begin{abstract}
This paper describes an approach for attractor selection (or multi-stability control) in nonlinear dynamical systems with constrained actuation. Attractor selection is obtained using two different deep reinforcement learning methods: 1)~the cross-entropy method (CEM) and 2)~the deep deterministic policy gradient (DDPG) method. The framework and algorithms for applying these control methods are presented. Experiments were performed on a Duffing oscillator, as it is a classic nonlinear dynamical system with multiple attractors. Both methods achieve attractor selection under various control constraints. While these methods have nearly identical success rates, the DDPG method has the advantages of a high learning rate, low performance variance, and a smooth control approach. This study demonstrates the ability of two reinforcement learning approaches to achieve constrained attractor selection.
\end{abstract}

\keywords{Coexisting attractors \and Attractor selection \and Reinforcement learning \and Machine learning \and Nonlinear dynamical system}

\section{Introduction}
Coexisting solutions or stable attractors are a hallmark of nonlinear systems and appear in highly disparate scientific applications~\cite{Thompson2002Nonlinear,Brun1985Observation,Maurer1980Effect,May1977Thresholds,Hudson1981Chaos,Wang2018Dynamics,Wang2019Nonlinear,Wang2020Nonlinear}. For these systems with multiple attractors, there often exists a preferable solution and one or more less preferable, or potentially catastrophic, solutions~\cite{Pisarchik2014Control}. For example, many nonlinear energy harvesters have multiple attractors, each with different levels of power output, among which the highest-power one is typically desired~\cite{Mann2009Energy,Stanton2010Nonlinear,Stanton2009Reversible}. Another example is the coexistence of period\nobreakdash-1 and period\nobreakdash-2 rhythms in cardiac dynamics. Controlling the trajectory of the cardiac rhythm to period\nobreakdash-1 is desirable to avoid arrhythmia~\cite{Kline1995Dynamical,Yehia1999Hysteresis}. Furthermore, coexisting solutions also appear in ecology and represent various degrees of ecosystem biodiversity~\cite{Scheffer2001Catastrophic}, where a bio-manipulation scheme is needed to avoid certain detrimental environmental states~\cite{Van2007Slow}. 

These and other applications have motivated the development of several control methods to switch between attractors of nonlinear dynamical systems. Pulse control is one of the simplest methods; it applies a specific type of perturbation to direct a system's trajectory from one basin of attraction to another and waits until the trajectory settles down to the desired attractor~\cite{Kaneko1989Chaotic,Kaneko1990Clustering,Samson1992Nonlinear}. Targeting algorithms, which were presented by Shinbrot et al.\ and modified by Macau et al., exploit the exponential sensitivity of basin boundaries in chaotic systems to small perturbations to direct the trajectory to a desired attractor in a short time~\cite{Shinbrot1990Using,Macau1999Driving}. Lai developed an algorithm to steer most trajectories to a desired attractor using small feedback control, which builds a hierarchy of paths towards the desirable attractor and then stabilizes a trajectory around one of the paths~\cite{Lai1996Driving}. Besides switching between naturally stable attractors, one can also stabilize the unstable periodic orbits and switch between these stabilized attractors. Since the OGY method was devised by Ott, Grebogi, and Yorke in 1990~\cite{Ott1996Controlling}, numerous works have built upon this original idea and explored relevant applications~\cite{Casas1997Control,Macau2006Control,Shinbrot1993Using,Yagasaki1998New,Epureanu1998Optimality,Hill2001Control}.

Although these methods can work for certain categories of problems, they are subject to at least one of the following restrictions: (1)~they only work for chaotic systems; (2)~they only work for autonomous systems; (3)~they need existence of unstable fixed points; (4)~they cannot apply constraints to control; or (5)~they cannot apply control optimization. Especially for the last two limitations, the compatibility with constrained optimal control is difficult to realize for most methods mentioned yet plays an important role in designing a real-world controller. For example, the limitations on the instantaneous power/force and total energy/impulse of a controller need be considered in practice. Another practical consideration is the optimization of total time and energy spent on the control process. Switching attractors using as little time or energy as possible is oftentimes required, especially when attempting to  escape detrimental responses or using a finite energy supply. 

Fortunately, a technique that is compatible with a broader scope of nonlinear systems, Reinforcement Learning (RL), can be applied without the aforementioned restrictions. By learning action-decisions while optimizing the long-term consequences of actions, RL can be viewed as an approach to optimal control of nonlinear systems~\cite{Chen1996RL}. Various control constraints can also be applied by carefully defining a reward function in RL~\cite{Sutton1992RL}. Although several studies of attractor selection using RL were published decades ago~\cite{Der1994Q,Gadaleta1999Optimal,Gadaleta2001Learning}, they dealt with only chaotic systems with unstable fixed points due to the limited choice of RL algorithms at the time. 

In recent years, a large number of advanced RL algorithms have been created to address complex control tasks. Most tasks with complicated dynamics have been implemented using the 3D physics simulator MuJoCo~\cite{Todorov2012Mujoco}, including control of Swimmer, Hopper, Walker~\cite{Levine2013Guided, Schulman2015Trust}, Half-Cheetah~\cite{Heess2015Learning}, Ant~\cite{Schulman2015High} and Humanoid~\cite{Tassa2012Synthesis} systems for balance maintenance and fast movement. Apart from simulated tasks, RL implementation in real-world applications includes motion planing of robotics~\cite{Mahmood2018Benchmarking}, autonomous driving~\cite{Lange2012Autonomous}, and active damping~\cite{Turner2020RL}. Furthermore, several researchers have explored RL-based optimal control for gene regulatory networks (GRNs), with the goal of driving gene expression towards a desirable attractor while using a minimum number of interventions~\cite{Datta2003External}. For example, Sirin et al.\ applied the model-free batch RL Least-Squares Fitted Q Iteration (FQI) method to obtain a controller from data without explicitly modeling the GRN~\cite{Sirin2013Employing}. Imani et al.\ used RL with Gaussian processes to achieve near-optimal infinite-horizon control of GRNs with uncertainty in both the interventions and measurements~\cite{Imani2017Control}. Papagiannis et al.\ introduced a novel learning approach to GRN control using Double Deep Q Network (Double DQN) with Prioritized Experience Replay~(PER) and demonstrated successful results for larger GRNs than previous approaches~\cite{ Papagiannis2019Deep, Papagiannis2019Learning}. Although these applications of RL for reaching GRNs' desirable attractors are related to our goal of switching attractors in continuous nonlinear dynamical systems, they are limited to Random Boolean Networks (RBN), which have discrete state and action spaces. A further investigation into generic nonlinear dynamical systems, where states and control inputs are oftentimes continuous, is therefore needed.

This paper will apply two RL algorithms, the cross-entropy method~(CEM) and deep deterministic policy gradient~(DDPG), to investigate the problem of attractor selection for a representative nonlinear dynamical system.

\section{Reinforcement Learning (RL) Framework}
\label{framework}

\begin{figure}
	\centering
	\includegraphics[width=0.6\linewidth]{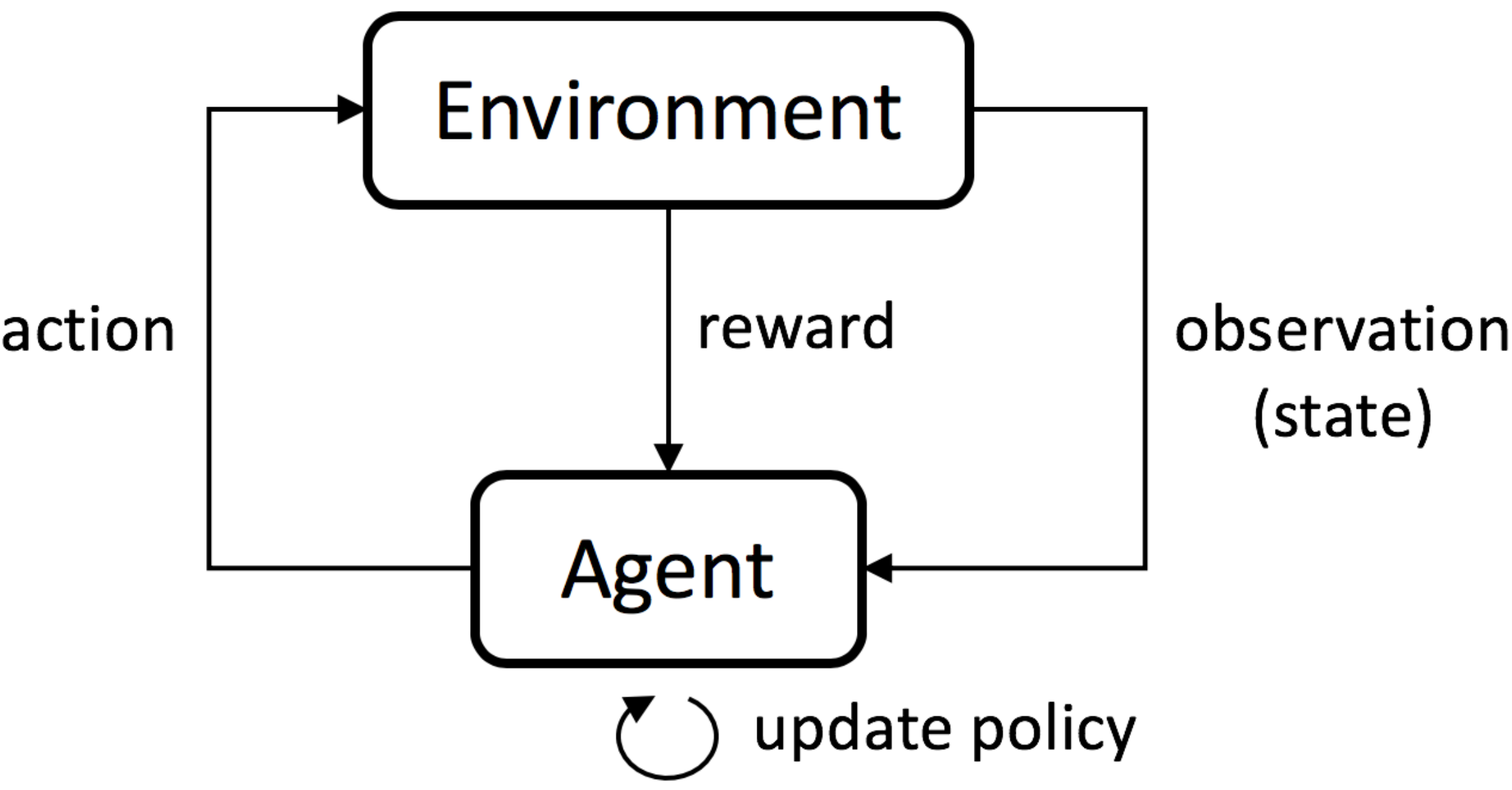}
	\caption{The typical framework of Reinforcement Learning. An agent has two tasks in each iteration: (1)~taking an action based on the observation from environment and the current policy; (2)~updating the current policy based on the immediate reward from environment and the estimated future rewards.}
	\label{fig:RL}
\end{figure}

In the RL framework shown in Fig.~\ref{fig:RL}, an \textit{agent} gains experience by making \textit{observations}, taking \textit{actions} and receiving \textit{rewards} from an \textit{environment}, and then learns a \textit{policy} from past experience to achieve goals (usually maximized cumulative \textit{reward}). 

To implement RL for control of dynamical systems, RL can be modeled as a Markov Decision Process (MDP):
\begin{enumerate}
	\item A set of environment states $S$ and agent actions $A$.
	\item $P_a(s, s') = \text{Pr}( s_{t+1} = s'\,|\, s_t = s, a_t=a )$ is the probability of transition from state $s$ at time $t$ to state $s'$ at time $t + 1$ with action $a$.
	\item $r_a (s,s')$ is the immediate reward after transition from $s$ to $s'$ with action $a$.
\end{enumerate}

A deterministic dynamical system under control is generally represented by the governing differential equation:
\begin{equation}
	\dot{x} = f(x,u,t),
\end{equation}
where $x$ is the system states, $\dot{x}$ is the states' rate of change, and $u$ is the control input. The governing equation can be integrated over time to predict the system's future states given initial conditions; it plays the same role as the transition probability in MDP. This equivalent interpretation of the governing equations and transition probability offers the opportunity to apply RL to control dynamical systems. The environment in RL can be either the system's governing equation if RL is implemented in simulation, or the experimental setup interacting with the real world if RL is implemented for a physical system. Instead of the conventional notation of states $x$ and control input $u$ in control theory, research in RL uses $s$ and $a$ to represent states and actions respectively. These RL-style notations are used throughout the remainder of this article. 

This paper implements RL algorithms to realize attractor selection (control of multi-stability) for nonlinear dynamical systems with constrained actuation. As a representative system possessing multiple attractors, the Duffing oscillator was chosen to demonstrate implementation details. 

\begin{figure*}
	\centering
	\includegraphics[width=0.9\linewidth]{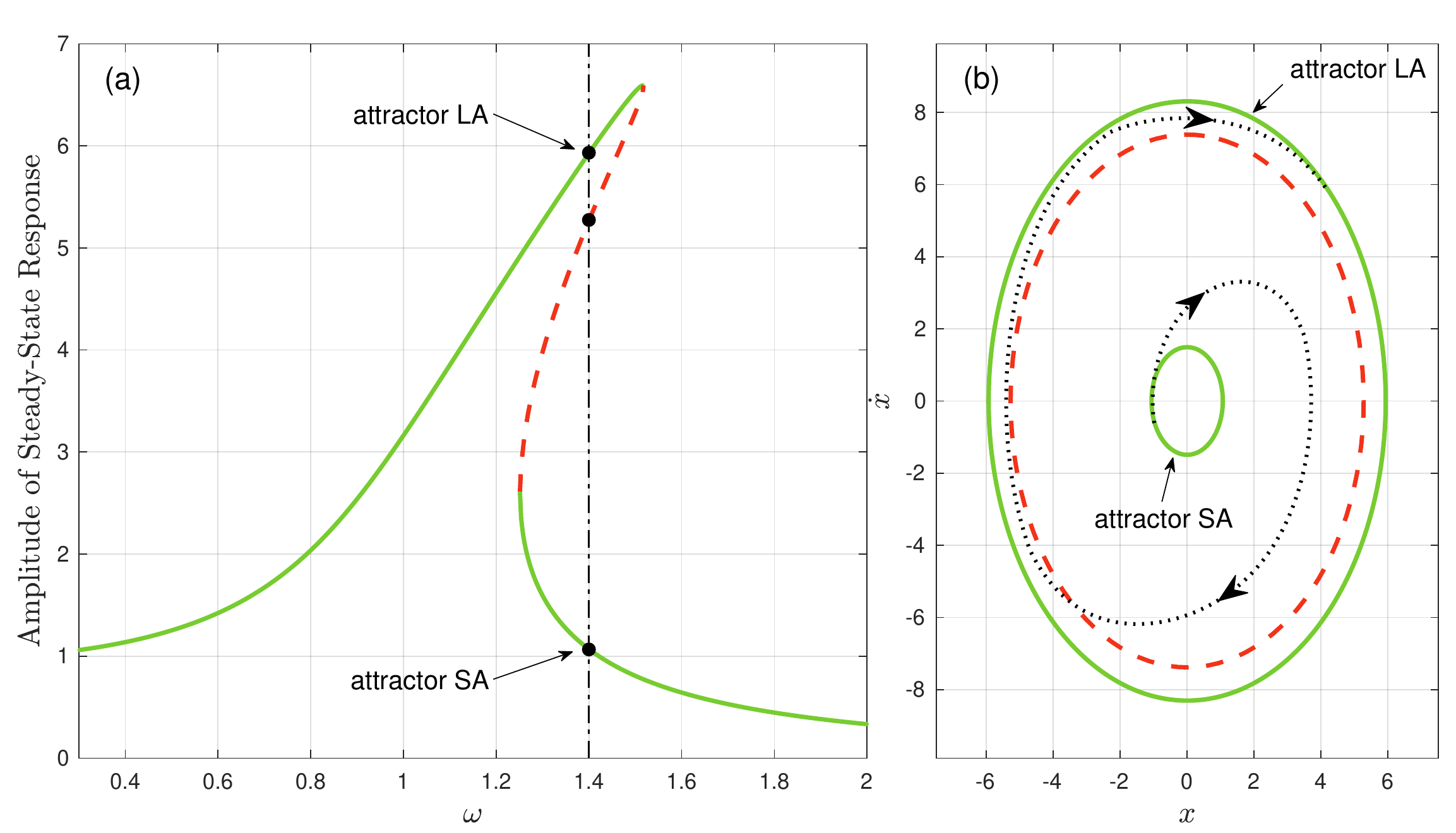}
	\caption{Coexisting attractors for the hardening Duffing oscillator: $\ddot{x} + 0.1 \dot{x} + x + 0.04 x^3 = \cos{\omega t}$. (a)~Frequency responses. (b)~Phase portrait of coexisting solutions when $\omega = 1.4$. Within a specific frequency range, periodic solutions coexist in (a), including two stable solutions (green solid lines) and an unstable solution (red dashed line) in-between. The stable solutions correspond to an attractor with small amplitude (SA) and one with large amplitude (LA). The dotted line in (b) is a trajectory of switching attractor SA $\rightarrow$ LA using the control method introduced in this paper.}
	\label{fig:Duffing}
\end{figure*}

\textbf{Environment}. A harmonically forced Duffing oscillator, which can be described by the equation $\ddot{x} + \delta \dot{x} + \alpha x + \beta x^3 = \Gamma \cos{\omega t}$, provides a familiar example for exploring the potential of RL for attractor selection. Fig.~\ref{fig:Duffing} shows an example of a hardening ($\alpha > 0$, $\beta > 0$) frequency response. For certain ranges of the parameters, the frequency response is a multiple-valued function of $\omega$, which represents multiple coexisting steady-state oscillations at a given forcing frequency. With a long-run time evolution, the oscillation of the unstable solution cannot be maintained due to inevitable small perturbations. The system will always eventually settle into one of the stable steady-state responses of the system, which are therefore considered ``attractors''. Our objective is to apply control to the Duffing oscillator to make it switch between the two attractors using constrained actuation. 

To provide actuation for attractor selection, an additional term $a(s)$, is introduced into the governing equation:
\begin{equation}
	\ddot{x} + \delta \dot{x} + \alpha x + \beta x^3 = \Gamma\cos{ \left( \omega t + \phi_0 \right) } + a(s).
	\label{eq:Duffing}
\end{equation}
where $a(s)$ is the actuation which depends on the system's states $s$. For example, if the Duffing oscillator is a mass--spring--damper system, $a(s)$ represents a force.

\textbf{Action}. Aligned with the practical consideration that an actuation is commonly constrained, the action term can be written as $a(s) := F \pi_\theta(s)$, where $F$ is the action bound which denotes the maximum absolute value of the actuation, and $\pi_\theta(s)$ is the control policy. $\pi_\theta(s)$ is designed to be a function parameterized by $\theta$, which has an input of the Duffing oscillator's states $s$, and an output of an actuation scaling value between $-1$ and $1$. Our objective is achieved by finding qualified parameters $\theta$ that cause the desired attractor to be selected.

\textbf{State \& Observation}. The Duffing oscillator is driven by a time-varying force $\Gamma\cos{\omega t}$; thus, the state should consist of position $x$, velocity $\dot{x}$, and time $t$. Given that $\Gamma\cos{\omega t}$ is a sinusoidal function with a phase periodically increasing from $0$ to $2\pi$, time can be replaced by phase for a simpler state expression. The system's state can therefore be written as $s := \left[ x,\, \dot{x},\, \phi \right]$, where $\phi$ is equal to $\omega t \text{ modulo } 2 \pi$. For the sake of simplicity, we have assumed that no observation noise was introduced and the states were fully observable by the agent. 

\textbf{Reward}. A well-trained policy should use a minimized control effort to reach the target attractor; thus the reward function, which is determined by state $s_t$ and action $a_t$, should inform the cost of the action taken and whether the Duffing oscillator reaches the target attractor. The action cost is defined as the impulse given by the actuation, $|{a}_t| \Delta t$, where $\Delta t$ is the time step size for control. The environment estimates whether the target attractor will be reached by evaluating the next state $s_{t+1}$. A constant reward of $r_\text{end}$ is given only if the target attractor will be reached.

\begin{figure*}
	\centering
	\includegraphics[width=0.9\linewidth]{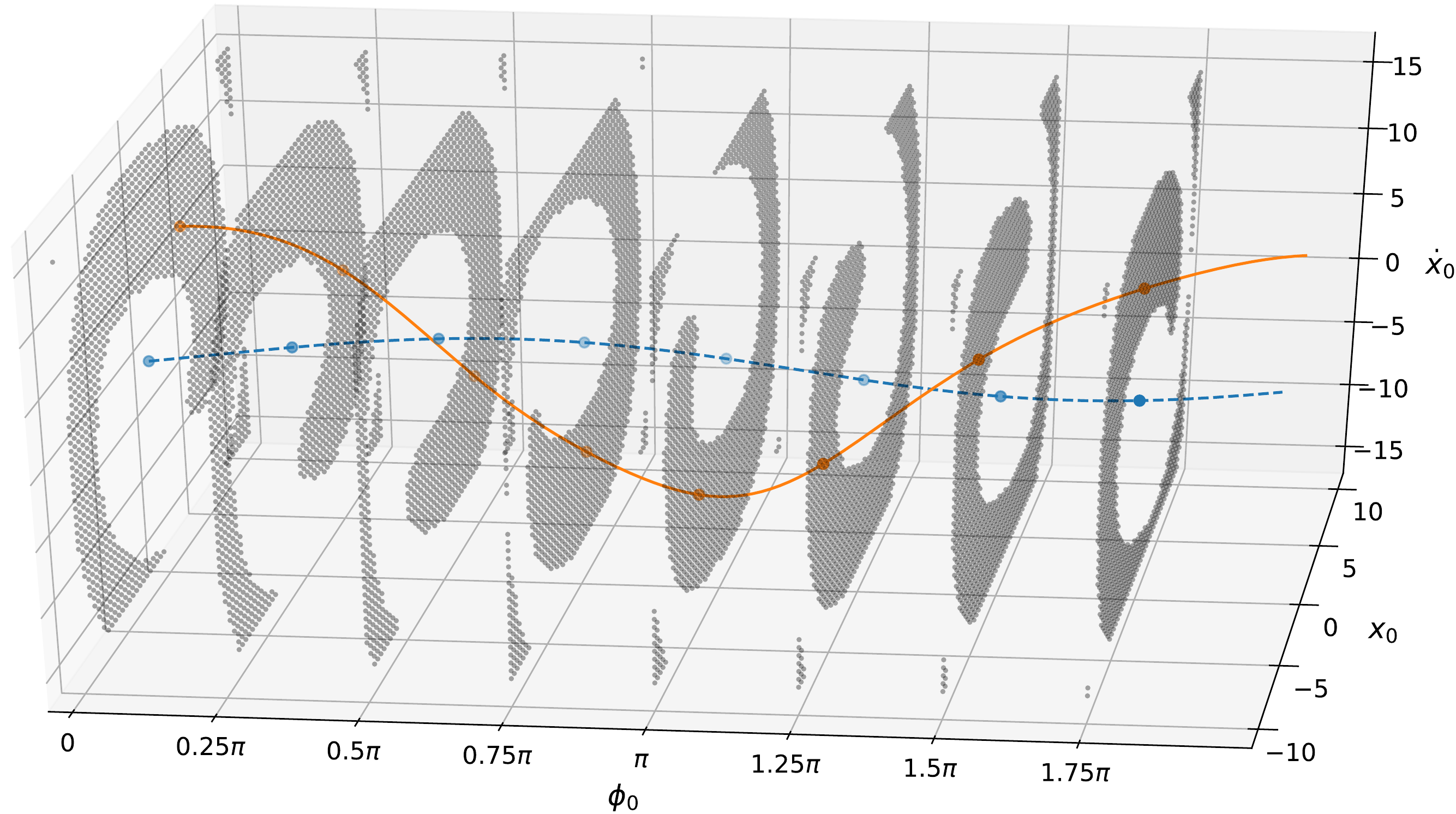}
	\caption{Basins of attraction (BoA) determined by the Duffing oscillator's state variables, $\left[ x,\, \dot{x},\, \phi \right]$. Each point in the BoA represents an initial condition which drives the system to a certain attractor without control. The orange solid line and the shaded areas are the large-amplitude attractor and its corresponding BoA, respectively. The blue dashed line and the blank areas are the small-amplitude attractor and its corresponding BoA, respectively.}
	\label{fig:BoA}
\end{figure*}

For estimating whether the target attractor will be reached, one could observe whether $s_{t+1}$ is in the ``basin of attraction'' of the target attractor. Basins of attraction (BoA) are the sets of initial states leading to their corresponding attractors as time evolves (see Fig.~\ref{fig:BoA}). Once the target BoA is reached, the system will automatically settle into the desired stable trajectory without further actuation. Therefore, the reward function can be written as:
\begin{equation}
	r(s_t, a_t) = -|a_t| \Delta t +
	\begin{cases}
		r_\text{end}, & \text{if $s_{t+1}$ reaches target BoA}\\
		0,     & \text{otherwise}
	\end{cases}
	\label{eq:reward}
\end{equation}

\textbf{BoA Classification}. Determining whether a state is in the target BoA is non-trivial. For an instantaneous state $s_t = [x_t,\, \dot{x}_t,\, \phi_t]$, we could set $a(s)=0$ in Eq.~\eqref{eq:Duffing} and integrate it with the initial condition $[x_0,\, \dot{x}_0,\, \phi_0] = s_t$. Integrating for a sufficiently long time should give a steady-state response, whose amplitude can be evaluated to determine the attractor. However, this prediction is needed for each time step of the reinforcement learning process, and the integration time should be sufficiently long to obtain steady-state responses; thus this approach results in expensive computational workload and a slow learning process. As a result, a more efficient method was needed for determining which attractor corresponded to the system's state~\cite{Wang2020Data}.

Since the number of attractors is finite, the attractor prediction problem can be considered a classification problem, where the input is the system's state and the output is the attractor label. Given that the boundary of the basins is nonlinear as shown in Fig.~\ref{fig:BoA}, the classification method of support vector machines~(SVM) with Gaussian kernel was selected for the Duffing oscillator. For other nonlinear dynamical systems, logistic regression is recommended for a linear boundary of basins, and a neural network is recommended for a large number of attractors. The training data was created by sampling states uniformly on the domain of three state variables, and the attractor label for each state was determined by the method mentioned above: evaluating future responses with long-term integration of governing equation. Generally speaking, this method transfers the recurring cost of integration during the reinforcement learning process to a one-time cost before the learning process begins. Collecting and labeling the data for training the classifier can be time consuming, but once the classifier is well-trained, the time for predicting the final attractor can be negligibly small. 

	\section{Algorithms}
\label{sec:algr}
This section describes two RL algorithms for attractor selection, the cross-entropy method (CEM) and deep deterministic policy gradient (DDPG). These two methods were selected for their ability to operate over continuous action and state spaces~\cite{Duan2016}. This section will first explain a few important terms before describing the specifics of each algorithm. 
\begin{enumerate}
	\item[] \textit{Phase 1}: the phase where the system is free of control, i.e., $a(s)=0$. The system is given a random initial condition at the beginning of Phase 1, waits for dissipation of the transient, and settles down on the initial attractor at the end of Phase 1.
	\item[] \textit{Phase 2}: the phase following Phase 1; the system is under control. Phase 2 starts with the system running in the initial attractor, and ends with either reaching the target BoA or exceeding the time limit. 
	\item[] \textit{Trajectory}: the system's time series for Phase 2. CEM uses trajectories of state-action pairs $[s_t,\, a_t]$, while DDPG uses trajectories of transitions $[s_t,\, a_t,\, r_t,\, s_{t+1}]$. The data of trajectories are stored in a replay buffer.
	\item[] \textit{Replay Buffer}: an implementation of experience replay~\cite{Lin1992Self}, which randomly selects previously experienced samples to update a control policy. Experience replay stabilizes the RL learning process and reduces the amount of experience required to learn~\cite{Mnih2015Human}.
\end{enumerate}
The entire process for learning a control policy can be summarized as iterative episodes. In each episode, the system runs through Phase 1 and Phase 2 in turn and the control policy is improved using the data of trajectories stored in the replay buffer. CEM and DDPG are different methods for updating the control policy.

\textbf{The cross-entropy method (CEM)} was pioneered as a Monte Carlo method for importance sampling, estimating probabilities of rare events in stochastic networks~\cite{Rubinstein1997, Rubinstein1999}. The CEM involves an iterative procedure of two steps: (1)~generate a random data sample according to a specified mechanism; and (2)~update the parameters of the random mechanism based on the data to produce a ``better'' sample in the next iteration~\cite{deBoer2005}.

In recent decades, an increasing number of applications of the CEM have been found for reinforcement learning~\cite{Mannor2003cross, Szita2006learning, Busoniu2009Policy}. CEM-based RL can be represented as an iterative algorithm~\cite{Lapan2018Deep}:
\begin{equation}
	\pi_{i+1} ( s ) = \mathop {\arg \min }\limits_{\pi_{i+1}} -\mathbb{E}_{z\sim\pi_{i} (s)} \left[ R(z) > \rho_i \right] \log \pi_{i+1}( s),
\end{equation}
where
\begin{equation}
	R(z) = \sum\limits_t r( s_t, a_t | z\sim\pi_{i} ( s ) ).
\end{equation}
$R(z)$ is the cumulative reward of a single time series trajectory generated by the current control policy $\pi_{i}(s)$, and $\rho_i$ is the reward threshold above which the trajectories are considered successful samples. This iteration minimizes the negative log likelihood (i.e., maximizes the log likelihood) of the most successful trajectories to improve the current policy. 

In our scenario where the control policy $\pi( s)$ is a neural network parameterized by $\theta$, the CEM iteration can also be written as:
\begin{equation}
	\theta_{i+1} = \mathop {\arg \min }\limits_{\theta} \frac{\sum\limits_j {L (a_j, F \pi_{\theta_{i}} (s_j ) ) \mathds{1}_A ( s_j ) }}{\sum\limits_j \mathds{1}_A ( s_j )},
\end{equation}
where
\begin{equation}
	A = \{s_j: s_j \in \mathcal{T}_k \,\,\text{and}\,\, R_{\mathcal{T}_k} > \rho_k\}.
\end{equation}
Given a state $s_j$ picked from the replay buffer, $L(\cdot,\cdot)$ is the loss function of the difference between its corresponding action value from past experience $a_j$, and the action value predicted by the current policy $F \pi_{\theta_{i}}$. The indicator function $\mathds{1}_A( s_j)$ has the value 1 when a state $s_j$ belongs to a successful trajectory $\mathcal{T}_k$ (i.e., the cumulative reward of the trajectory $R_{\mathcal{T}_k}$ is greater than the a threshold $\rho_k$), and has the value 0 otherwise. The detailed CEM algorithm designed for attractor selection is presented in the experiment section.	

\textbf{Deep Deterministic Policy Gradient (DDPG)} is a RL algorithm that can operate over continuous state and action spaces~\cite{Lillicrap2015DDPG}. The goal of DDPG is to learn a policy which maximizes the expected return $J = \mathbb{E}_{r_i, s_i, a_i}[R_{t=1}]$, where the return from a state is defined as the sum of discounted future rewards $R_t = \sum\nolimits_{i = t}^T \gamma^{i-t} r(s_i, a_i)$ with a discount factor $\gamma \in [0,\,1]$.

An action-value function, also known as a ``critic'' or ``Q-value'' in DDPG, is used to describe the expected return after taking an action $a_t$ in state $s_t$:
\begin{equation}
	Q(s_t, a_t) = \mathbb{E}_{r_{i \geqslant t}, s_{i>t}, a_{i>t}}[R_t | s_t,a_t].
\end{equation}
DDPG uses a neural network parameterized by $\psi$ as a function appropriator of the critic $Q(s, a)$, and updates this critic by minimizing the loss function of the difference between the ``true'' Q-value $Q(s_t, a_t)$ and the ``estimated'' Q-value $y_t$:
\begin{equation}
	L( \psi ) = \mathbb{E}_{s_t, a_t, r_t} \left[ \left( Q( s_t, a_t | \psi) - y_t \right) ^ 2 \right],
\end{equation} 
where
\begin{equation}
	y_t = r(s_t, a_t) + \gamma Q( s_{t+1}, \pi( s_{t+1} )_t | \psi).
	\label{eq:y_t}
\end{equation}
Apart from the ``critic'', DDPG also maintains an ``actor'' function to map states to a specific action, which is essentially our policy function $\pi(s)$. DDPG uses another neural network parameterized by $\theta$ as a function approximator of the actor $\pi(s)$, and updates this actor using the gradient of the expected return $J$ with respect to the actor parameters $\theta$:
\begin{equation}
	\nabla_\theta J \approx \mathbb{E}_{s_t}\left[ \nabla_a Q_\psi(s, a) |_{s=s_t, a=\pi(s_t)} \nabla_\theta \pi_\theta (s) |_{s=s_t} \right].
	\label{dJ}
\end{equation}	

In order to enhance the stability of learning, DDPG is inspired by the success of Deep Q Network (DQN)~\cite{Mnih2013Playing, Mnih2015Human} and uses a ``replay buffer'' and separate ``target networks'' for calculating the estimated Q-value $y_t$ in Eq.~\eqref{eq:y_t}. The replay buffer stores transitions $[s_t,\, a_t,\, r_t,\, s_{t+1}]$ from experienced trajectories. The actor $\pi(s)$ and critic $Q(s, a)$ are updated by randomly sampling a minibatch from the buffer, allowing the RL algorithm to benefit from stably learning across uncorrelated transitions. The target networks are copies of actor and critic networks, $\pi'_{\theta'}(s)$ and $Q'_{\psi'}(s, a)$ respectively, that are used for calculating the estimated Q-value $y_t$. The parameters of these target networks are updated by slowly tracking the learned networks $\pi_{\theta}(s)$ and $Q_{\psi}(s, a)$:
\begin{equation}
	\begin{split}
		&\psi' \leftarrow \tau \psi + (1-\tau) \psi', \\
		&\theta' \leftarrow \tau \theta + (1-\tau) \theta',
	\end{split}
\end{equation}
where $0 < \tau \ll 1$. This soft update constrains the estimated Q-value $y_t$ to change slowly, thus greatly enhancing the stability of learning. The detailed DDPG algorithm designed for attractor selection is presented in the experiment section.

One major difference between CEM and DDPG is the usage of the policy gradient $\nabla_\theta J$. DDPG computes the gradient for policy updates in Eq.\eqref{dJ} while CEM is a gradient-free algorithm. Another difference lies in their approach to using stored experience in the replay buffer. CEM improves the policy after collecting new trajectories of data, while DDPG continuously improves the policy at each time step as it explores the environment~\cite{Duan2016}.

\setcounter{magicrownumbers}{0}
\begin{table*}
	\caption{Algorithm: Cross-Entropy Method (CEM) for Attractor Selection}
	\label{table:CEM}
	\begin{tabular}{r|ll}
		\toprule
		\rownumber & Randomly initialize policy network ${\pi _\theta }(s)$ with weight $\theta$ &\\
		\rownumber & Set the initial condition of the Duffing equation $s_0 = [x_0,\, \dot{x}_0,\, \phi_0]$ &\\
		\rownumber & Set time of Phase 1 and Phase 2, $T_1$ and $T_2$ &\\
		\rownumber & Set best sample proportion, $p \in [0,\,1]$ &\\
		\rownumber & \textbf{for} episode = 1 : M \textbf{do} &\\
		\rownumber & \quad Initialize an empty replay buffer $B$ &\\
		\rownumber & \quad \textbf{for} sample = 1 : N \textbf{do}&\\
		\rownumber & \quad\quad Initialize an empty buffer $\tilde{B}$ &\\
		\rownumber & \quad\quad Initialize a random process $\mathcal{N}$ for action exploration &\\
		\rownumber & \quad\quad Initialize trajectory reward, $R = 0$ &\\
		\rownumber & \quad\quad Add noise to time of Phase 1, ${T_1}'=T_1+\text{random}(0, 2\pi / \omega)$ & \\
		\rownumber & \quad\quad Integrate Eq.~\eqref{eq:Duffing} for $t \in \left[0,\, {T_1}' \right]$ with $a(s)=0$ & Phase 1\\
		\rownumber & \quad\quad \textbf{for} $t = {T_1}'$ : ${T_1}' + T_2$ \textbf{do} & Phase 2\\
		\rownumber & \quad\quad\quad Observe current state, $s_t = \left[ x_t,\, \dot{x}_t,\, \phi_t \right]$ &\\
		\rownumber & \quad\quad\quad Evaluate action $a_t(s_t) = F\pi_\theta(s_t) + \mathcal{N}_t$, according to the current policy and exploration noise &\\
		\rownumber & \quad\quad\quad Step into the next state $s_{t+1}$, by integrating Eq.~\eqref{eq:Duffing} for $\Delta t$ &\\
		\rownumber & \quad\quad\quad Update trajectory reward $R \leftarrow R + r(s_t, a_t)$, according to Eq.~\eqref{eq:reward} &\\
		\rownumber & \quad\quad\quad Store state-action pair $[s_t, a_t]$ in $\tilde{B}$ &\\
		\rownumber & \quad\quad\quad Evaluate the basin of attraction for $s_{t+1}$ &\\
		\rownumber & \quad\quad\quad \textbf{if} the target attractor's basin is reached: &\\
		\rownumber & \quad\quad\quad\quad Label each state-action pair $[s_t, a_t]$ in $\tilde{B}$ with trajectory reward $R$, and append them to $B$ &\\
		\rownumber & \quad\quad\quad\quad \textbf{break} &\\
		\rownumber & \quad\quad \textbf{end for} &\\
		\rownumber & \quad \textbf{end for} &\\
		\rownumber & \quad Choose the a minibatch of the elite $p$ proportion of $(s, a)$ in $B$ with the largest reward $R$ &\\
		\rownumber & \quad Update policy ${\pi _\theta }(s)$ by minimizing the loss, $L = \frac{1}{\text{Minibatch Size}}\sum\limits_i {\left( F \pi_\theta(s_i) - a_i \right)^2 }$ &\\
		\rownumber & \textbf{end for} &\\
		\bottomrule
	\end{tabular}
\end{table*}

\setcounter{magicrownumbers}{0}
\begin{table*}
	\caption{Algorithm: Deep Deterministic Policy Gradient (DDPG) for Attractor Selection}
	\label{table:DDPG}
	\begin{tabular}{r|ll}
		\toprule
		\rownumber & Randomly initialize actor network ${\pi _\theta }(s)$ and critic network $Q_\psi(s, a)$ with weights $\theta$ and $\psi$ &\\
		\rownumber & Initialize target network $\pi'_{\theta'}(s)$ and $Q'_{\psi'}(s, a)$ with weights $\theta' \leftarrow \theta$, $\psi' \leftarrow \psi$ &\\
		\rownumber & Set the initial condition of the Duffing equation $s_0 = [x_0,\, \dot{x}_0,\, \phi_0]$ &\\
		\rownumber & Set discount factor $\gamma$, and soft update factor $\tau$ &\\
		\rownumber & Set time of Phase 1 and Phase 2, $T_1$ and $T_2$ &\\
		\rownumber & Initialize replay buffer $B$ &\\
		\rownumber & \textbf{for} episode = 1 : M \textbf{do} &\\
		\rownumber & \quad Initialize a random process $\mathcal{N}$ for action exploration &\\
		\rownumber & \quad Add noise to time of Phase 1, ${T_1}'=T_1+\text{random}(0, 2\pi / \omega)$ &\\
		\rownumber & \quad Integrate Eq.~\eqref{eq:Duffing} for $t \in \left[0,\, {T_1}' \right]$ with $a(s)=0$ & Phase 1\\
		\rownumber & \quad\textbf{for} $t = {T_1}'$ : ${T_1}' + T_2$ \textbf{do} & Phase 2\\
		\rownumber & \quad\quad Observe current state, $s_t = \left[ x_t,\, \dot{x}_t,\, \phi_t \right]$ &\\
		\rownumber & \quad\quad Evaluate action $a_t(s_t) = F\pi_\theta(s_t) + \mathcal{N}_t$, according to the current policy and exploration noise &\\
		\rownumber & \quad\quad Step into the next state $s_{t+1}$, by integrating Eq.~\eqref{eq:Duffing} for $\Delta t$ &\\
		\rownumber & \quad\quad Evaluate reward $r_t(s_t, a_t)$, according to Eq.~\eqref{eq:reward} &\\
		\rownumber & \quad\quad Store transition $[s_t,\, a_t,\, r_t,\, s_{t+1}]$ in $B$ &\\
		\rownumber & \quad\quad Sample a random minibatch of $N$ transitions $[s_i,\, a_i,\, r_i,\, s_{i+1}]$ from $B$ &\\
		\rownumber & \quad\quad Set $y_i = r_i + \gamma Q'_{\psi'}(s_{i+1}, F\pi'_{\theta'}(s_{i+1}))$ &\\
		\rownumber & \quad\quad \makecell[l]{Update the critic network by minimizing the loss: \\ \quad\quad\quad\quad $L=\frac{1}{N}\sum\limits_i (y_i - Q_\psi(s_i, a_i))^2$} &\\
		\rownumber & \quad\quad \makecell[l]{Update the actor network using the sampled policy gradient: \\ \quad\quad\quad\quad $\nabla_\theta J \approx \frac{1}{N}\sum\limits_i \nabla_a Q_\psi(s, a) |_{s=s_i, a=F\pi_\theta(s_i)} \nabla_\theta \pi_\theta(s)|_{s=s_i}$   } &\\
		\rownumber & \quad\quad \makecell[l]{Update the target networks: \\ \quad\quad\quad\quad $\psi' \leftarrow \tau \psi + (1-\tau) \psi',\, \theta' \leftarrow \tau \theta + (1-\tau) \theta'$} &\\
		\rownumber & \quad\quad \textbf{if} the target attractor's basin is reached in $s_{t+1}$: \textbf{break}&\\
		\rownumber & \quad \textbf{end for} &\\
		\rownumber & \textbf{end for} & \\
		\bottomrule
	\end{tabular}
\end{table*}

\section{Experiment}
This section presents the details of the experiment performing attractor selection for the Duffing oscillator using CEM and DDPG. 

The governing equation for the Duffing oscillator is given by Eq.~\eqref{eq:Duffing}, where $\delta = 0.1$, $\alpha = 1$, $\beta = 0.04$, $\Gamma = 1$ and $\omega = 1.4$. The Duffing equation is integrated using scipy.integrate.odeint() in Python with a time step of 0.01. The time step for control is 0.25, and reaching the target BoA will add a reward $r_\text{end} = 100$. Therefore, the reward function Eq.~\eqref{eq:reward} can be written as:
\begin{equation}
	r(s_t, a_t)= 
	\begin{cases}
		100 - 0.25 |a_t|, & \text{if $s_{t+1}$ reaches target BoA,}\\
		- 0.25 |a_t|,     & \text{otherwise.}
	\end{cases}
	\label{eq:reward_SM}
\end{equation}
For the estimation of BoA, the SVM classifier with radial basis function (RBF) kernel was trained using a \(50\times50\times50\) grid of initial conditions, with $x \in [-10,\, 10]$, $\dot{x} \in [-15,\, 15]$ and $\phi \in [0,\, 2\pi]$.

The detailed attractor selection algorithm using CEM can be found in Tab.~\ref{table:CEM}. In line 11, the time of Phase 1 is perturbed by adding a random value between $0$ and $2\pi/\omega$ (a forcing period). This noise provides diversity of the system's states at the beginning of Phase 2, which enhances generality and helps prevent over-fitting of the control policy network $\pi_\theta(s)$. The policy network $\pi_\theta(s)$ has two fully connected hidden layers, each of which has 64 units and an activation function of ReLU \cite{Glorot2011Deep}. The final output layer is a tanh layer to bound the actions. Adam optimizer \cite{Kingma2014Adam} with a learning rate of $10^{-3}$ and a minibatch size of 128 was used for learning the neural network parameters. For the system's settling down in Phase 1 we used $T_1 = 15$, and for constraining the time length of control we used $T_2 = 20$. In each training episode, state-action pairs from 30 trajectory samples were stored in the replay buffer ($N = 30$), and those with reward in the top 80\% were selected for training the network ($p = 0.8$). 

The detailed attractor selection algorithm using DDPG can be found in Tab.~\ref{table:DDPG}. Apart from the ``actor'' network $\pi_\theta(s)$ which is same as the policy network in the CEM, the DDPG introduces an additional ``critic'' network $Q_\psi(s, a)$. This $Q$ network is designed to be a neural network parameterized by $\psi$, which has the system's state and corresponding action as inputs, and a scalar as the output. As in the algorithm using CEM, the state diversity is promoted by introducing noise to the time of Phase 1 in line 9. Both the actor network $\pi_\theta(s)$ and the critic network $Q_\psi(s, a)$ have two hidden layers, each of which has 128 units and an activation function of ReLU \cite{Glorot2011Deep}. For the actor network, the final output layer is a tanh layer to bound the actions. For the critic network, the input layer consists of only the state $s$, while the action $a$ is included in the 2nd hidden layer. Adam optimizer \cite{Kingma2014Adam} was used to learn the neural network parameters with a learning rate of $\tau_\theta = 10^{-4}$ and $\tau_\psi = 10^{-3}$ for the actor and critic respectively. For the update of the critic network we used a discount factor of $\gamma = 0.9$. For the soft update of the target network $\pi'_{\theta'}(s)$ and $Q'_{\psi'}(s, a)$ by Polyak Averaging, we used $\tau = 0.1$. For the system's settling down in Phase 1 we used $T_1 = 15$, and for constraining the time length of control we used $T_2 = 20$. The replay buffer had a size of $10^6$. In each episode, the minibatch of transitions sampled from the replay buffer had a size of $N=64$.

To test the CEM and DDPG algorithms, constraints were constructed with varying levels of difficulty, i.e., different action bounds $F$. Recall that the control term in Eq.~\eqref{eq:Duffing} can be written as $a(s) = F\pi_\theta(s)$, which is bounded between $-F$ and $F$. It's also worth noting that each policy only controls a one-way trip of attractor switching. For the example of a Duffing oscillator with two attractors, one control policy is needed for transitioning from the small-amplitude attractor to the large-amplitude one, and another control policy is needed for the reverse direction. 

\begin{figure*}
	\centering
	\includegraphics[width=\linewidth]{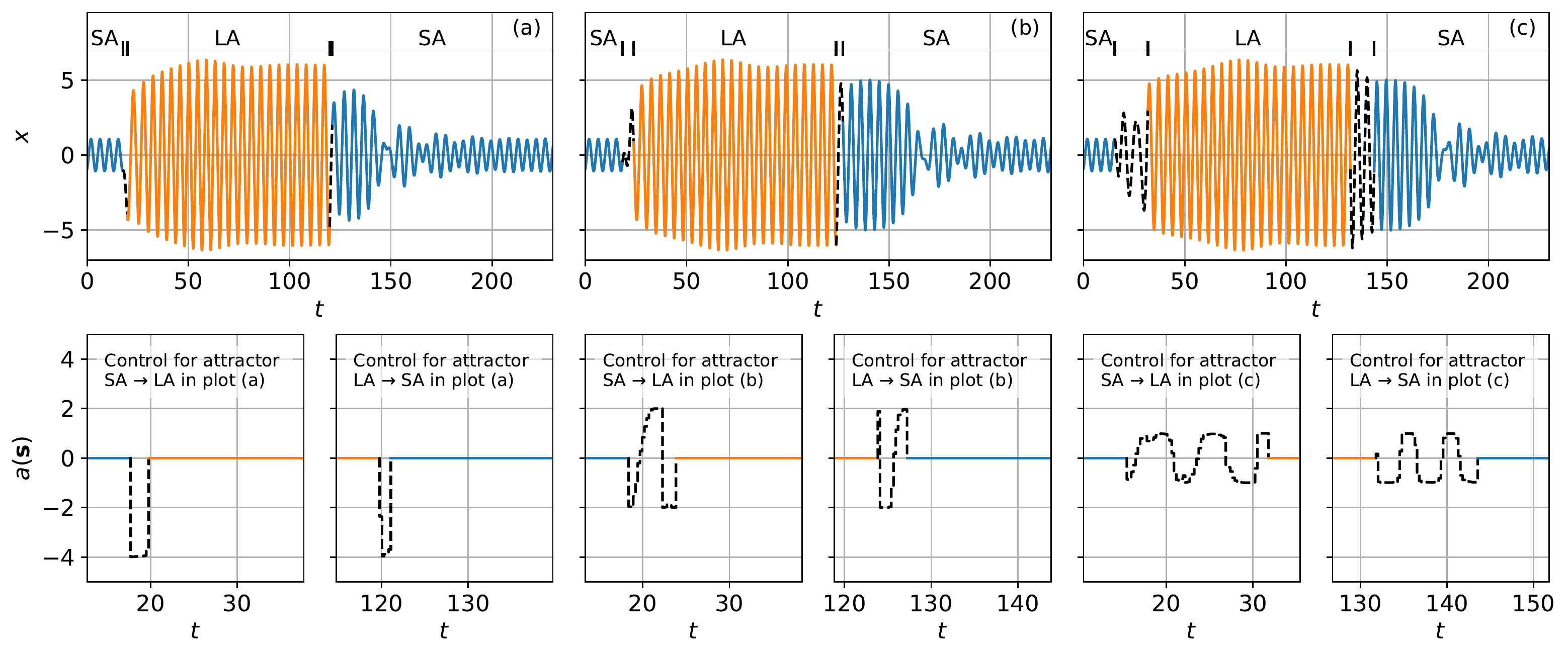}
	\caption{Attractor selection using control policy learned by CEM with varying action bounds: (a) $F = 4$, (b) $F = 2$, (c) $F = 1$. The blue lines represent the Duffing oscillator running in the basin of attractor SA, which has a periodic solution with small amplitude. The orange lines represent the Duffing oscillator running in the basin of attractor LA, which has a periodic solution with large amplitude. The black dashed lines represent the Duffing oscillator under control. Each plot of the system's responses in the first row corresponds to the two sub-plots of the control processes in the second row: attractor SA $\rightarrow$ LA and attractor LA $\rightarrow$ SA.}
	\label{fig:CEM}
\end{figure*}
\begin{figure*}
	\centering
	\includegraphics[width=\linewidth]{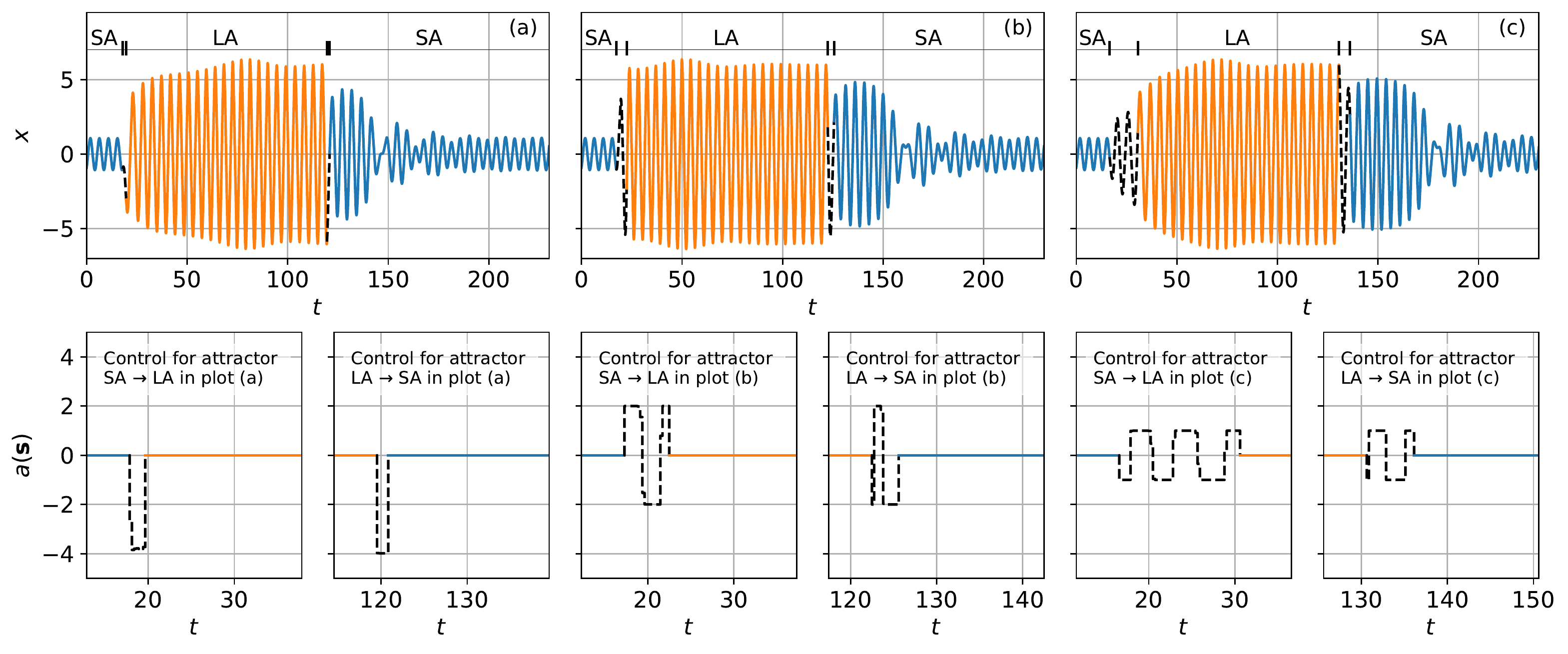}
	\caption{Attractor selection using control policy learned by DDPG with varying action bounds: (a) $F = 4$, (b) $F = 2$, (c) $F = 1$. The blue lines represent the Duffing oscillator running in the basin of attractor SA, which has a periodic solution with small amplitude. The orange lines represent the Duffing oscillator running in the basin of attractor LA, which has a periodic solution with large amplitude. The black dashed lines represent the Duffing oscillator under control. Each plot of the system's responses in the first row corresponds to the two sub-plots of the control processes in the second row: attractor SA $\rightarrow$ LA and attractor LA $\rightarrow$ SA.}
	\label{fig:DDPG}
\end{figure*}

\section{Results}
\label{sec:result}

\begin{figure*}
	\centering
	\includegraphics[width=0.7\linewidth]{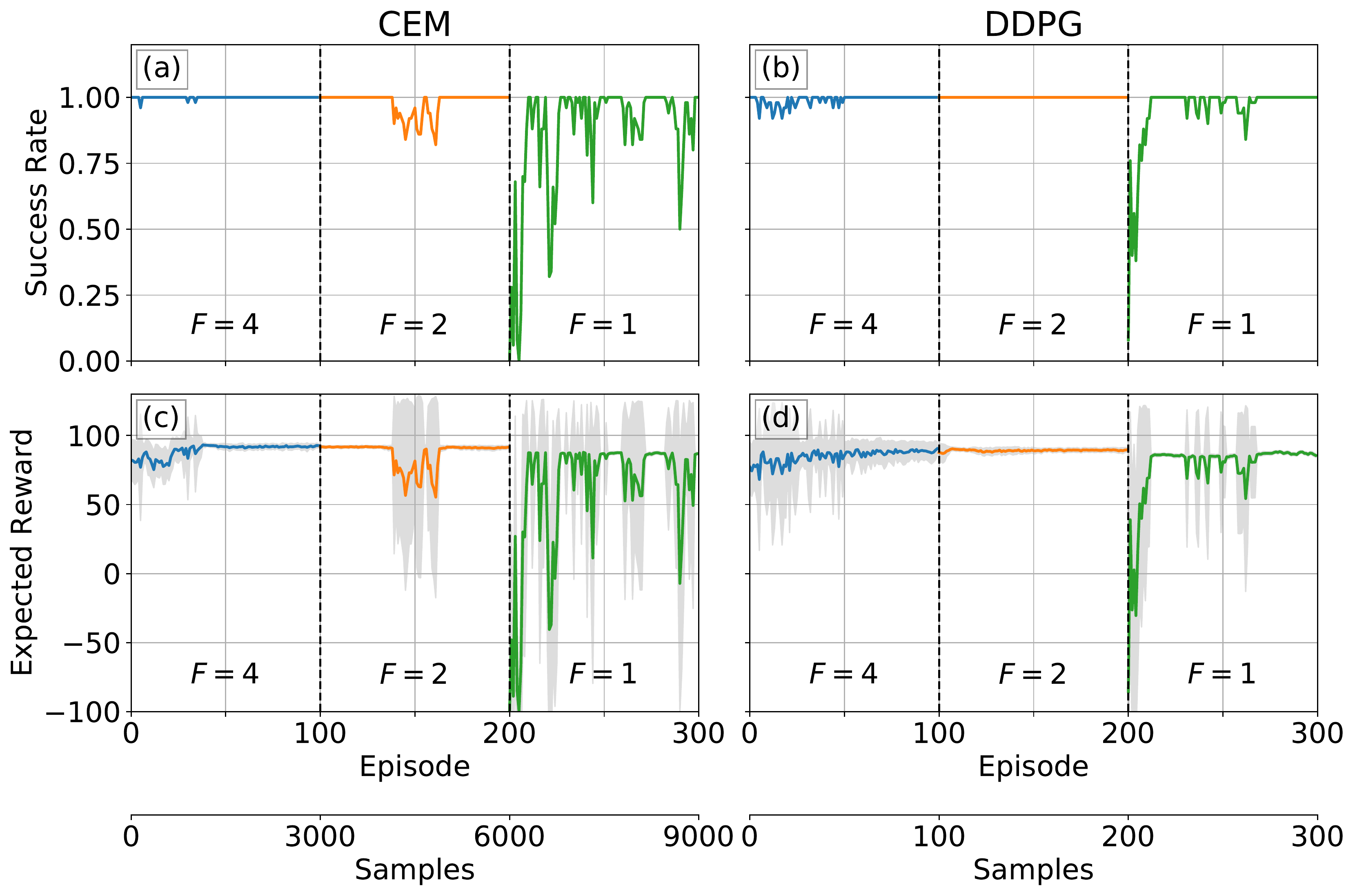}
	\caption{The control policy performance curve for the success rate of reaching target attractor using (a) CEM and (b) DDPG, and the expected accumulated reward of a control process using (c) CEM and (d) DDPG. The varying action bound of 4, 2, and 1 are represented by the blue, orange, and green lines respectively. The gray areas in plot (c) and (d) represent standard deviation of the accumulated rewards.}
	\label{fig:learning_curve}
\end{figure*}

Fig.~\ref{fig:CEM} and Fig.~\ref{fig:DDPG} show the time series of attractor selection using the control policy learned by the CEM and DDPG algorithms, respectively. For simplicity, the attractor with a small amplitude of steady-state response is named ``SA'' while that with a large amplitude is named ``LA''.

For all six test cases in Fig.~\ref{fig:CEM} and Fig.~\ref{fig:DDPG}, the Duffing oscillator starts near the SA solution, then switches from SA to LA, and finally switches backwards from LA to SA. Compared with the short time length for control (region of black lines), the attractor selection spends more time waiting for dissipation of the transient process, where the system is automatically approaching the target attractor under no control. This observation shows a smart and efficient strategy the control policy has learned: instead of driving the system precisely to the state set of the target attractor, it just drives the system to the attractor's basin, where the system might be far away from the target attractor initially but will reach it without further control effort as time evolves.

One can also observe that a smaller action bound results in a longer time length of control for both CEM and DDPG algorithms. It can be qualitatively explained by the energy threshold for jumping from one attractor to another. The energy provided by the actuation should be accumulated beyond the energy threshold to push the system away from one attractor. A smaller action bound therefore leads to longer time for the energy accumulation. 

Another observation is that the system quickly approaches near LA when starting from SA, while it takes more time to approach near SA when starting from LA. This can be explained using the unstable solution of the Duffing equation which is represented as the red dashed circle in Fig.~\ref{fig:Duffing}~(b). This circle can be considered the boundary between the basins of two attractors, across which the system jumps from one attractor to another. This boundary is close to LA, which means that the system will be near LA immediately after going across the boundary from SA. In contrast, SA is far from the boundary; therefore, a longer time is needed to reach near SA after going across the boundary from LA. The attractors' distances from the unstable solution also indicate their robustness and likelihood of occurrence. In this Duffing oscillator, SA is more robust and more likely to occur than LA. The successful switching from a more likely attractor (such as SA) to a less likely attractor (such as LA), which is difficult for traditional control methods, is another advantage of the proposed RL-based methods.

Although the CEM and DDPG algorithms both achieve attractor selection with various action bounds, DDPG has advantages of providing smooth actuations. From the comparison of the actuation $F\pi_\theta(s)$ between Fig.~\ref{fig:CEM} and Fig.~\ref{fig:DDPG}, the actuation from CEM shows more jagged motion than that from DDPG especially for the small action bound. More advantages of DDPG over CEM can be observed in Fig.~\ref{fig:learning_curve}, which compares the trend of their policies' success rate and expected reward during learning process. In each learning episode, 100 samples with random initial states were used to evaluate the control policy's success rate of reaching the target attractor and the mean and standard deviation of the samples' accumulated reward. For faster convergence, the policy trained for a tighter action bound was initialized with the parameters of a well-trained policy for a more generous action bound. In other words, instead of learning from nothing, a policy learned to solve a hard problem from the experience of solving an easy one. The ``untrained'' policy for the action bound of 2 was initialized with the parameters of the ``trained'' policy for the action bound of 4, and so on. Therefore in Fig.~\ref{fig:learning_curve}, the action bound decreases as the learning episode increases. 

The success rates and expected rewards were equally high for both CEM and DDPG at the end of the learning for each action bound (episode = 100, 200, 300), but DDPG converged faster, especially for the action bound of 1. This advantage of DDPG can be observed from the performance curve which evolves with ``episode'' in Fig.~\ref{fig:learning_curve}, but DDPG is even better than this considering that it spends much less time on each episode than CEM. As shown in the Table~\ref{table:CEM}, CEM needs $N$ samples of trajectories for updating the policy in each episode, while DDPG (see Table~\ref{table:DDPG}) collects only the sample of a single trajectory. The real learning speed can therefore be reflected by the total ``samples'' generated instead of ``episode''. Fig.~\ref{fig:learning_curve} shows the DDPG's advantage of learning speed and data efficiency by providing an additional horizontal axis for the number of samples, where CEM generates 30 samples per episode while DDPG generates only 1 sample per episode. Parallel computing can be used for helping CEM narrow the gap, but generally DDPG has a natural advantage of learning speed. Furthermore, after the CEM learned an optimal policy, it often degenerated to a sub-optimal policy, which can be observed from the small perturbation around episode 150 and the significant oscillation throughout episode 200--300 in Fig.~\ref{fig:learning_curve}~(b,d). In contrast, DDPG shows a lower variance of the policy performance after an optimal policy has been learned, which can be observed from the flat line in episode 100--200 and the comparatively small perturbation in episode 200--300. 

\section{Conclusion}
\label{sec:conclusion}

This work applies advanced reinforcement learning (RL) methods to the control problem of attractor selection, resulting in algorithms that switch between coexisting attractors of a nonlinear dynamical system using constrained actuation. A control framework was presented combining attractor selection and general RL methods. Two specific algorithms were designed based on the cross-entropy method (CEM) and deep deterministic policy gradient (DDPG). 

The Duffing oscillator, which is a classic nonlinear dynamical system with multiple coexisting attractors, was selected to demonstrate the control design and show the performance of the control policy learned by CEM and DDPG. Both methods achieved attractor selection under various constraints on actuation. They had equally high success rates, but DDPG had advantages of smooth control, high learning speed, and low performance variance. 

The RL-based attractor selection is model-free and thus needs no prior knowledge on the system model. This provides broad applicability since precise mathematical models for many real-world systems are often unavailable due to their highly nonlinear features or high noise level. Additionally, the control constraints can be easily customized for a wide variety of problems. Apart from constraining the maximum actuation in the Duffing oscillator example, the system's position can be constrained in case of working in a limited space, the system's velocity can be constrained if there exists a hardware requirement of speed range, and the constraints themselves can even be time-varying. Various constraints can be realized by carefully designing the action term and reward function. 

Future work needs to extend our investigations in three key directions. First, although the proposed approach is model-free and does not require a priori knowledge of the system dynamics when training the control policy, obtaining the basins of attraction (BoAs) might still require prior knowledge of the system's equilibria and stability behavior as derived from the governing equation. In order to entirely get rid of this model dependence, a data-driven approach to automatically finding coexisting attractors based on simulation or experimental data will be needed. Second, the quantity of samples for training a qualified BoA classifier needs to be reduced. As mentioned in the experiment section, the classifier was trained using a $50\times50\times50$ grid of 3-dimensional initial conditions. Each of $125,000$ samples was obtained by running a simulation and evaluating its final state. A large number of training samples will become a huge burden when (1)~simulations are computationally expensive, (2)~samples are collected from experiment, or (3)~the dimension of initial conditions is high. A more data-efficient sampling method is therefore needed. Third, we would like to reiterate that the model-free approach proposed in this paper only indicates that a model is unnecessary for learning a control policy, but does not mean the model is useless. If the approach is implemented in real-world experiments which are much more costly than simulations, a more efficient way is to first find a sub-optimal control policy in simulation and then ``transfer'' the pre-trained policy to the experiments for further optimization. This process is called ``transfer learning'', where the heavy-learning workload in experiments is shared with simulations, and the simulation-based RL will need a model as its environment. More studies implementing the attractor selection approach based on real-world experiments and transfer learning are certainly worthy topics for further investigations.

This study demonstrated two reinforcement learning approaches for constrained attractor selection. These approaches make it possible to switch between attractors without the limitations of prior methods in the literature. By optimizing the policy subject to realistic constraints, they are applicable to a wide variety of practical problems.

\bibliographystyle{unsrt}  
\bibliography{references}  

%
%
%
%

\end{document}